# Integrated superconducting detectors on semiconductors for quantum optics applications

M. Kaniber,[1] F. Flassig,[1] G. Reithmaier,[1] R. Gross,[2,3] and J. J. Finley[1,3]

[1] Walter Schottky Institut and Physik Department, Technische Universität München, Am Coulombwall 4, 85748 Garching, Germany
[2] Walther-Meißner-Institut, Bayerische Akademie der Wissenschaften und Physik-Department, Technische Universität München, 85748 Garching, Germany
[3] Nanosystems Initiative Munich, Schellingstraße 4, 85748 München, Germany

**Abstract:**

Semiconductor quantum photonic circuits can be used to efficiently generate, manipulate, route and exploit non-classical states of light for distributed photon based quantum information technologies. In this article, we review our recent achievements on the growth, nanofabrication and integration of high-quality, superconducting Niobium nitride thin films on optically active, semiconducting GaAs substrates and their patterning to realise highly efficient and ultrafast superconducting detectors on semiconductor nanomaterials containing quantum dots. Our state-of-the-art detectors reach external detection quantum efficiencies up to $20\,\%$ for $\sim 4\,nm$ thin films and single photon timing resolutions $< 72\,ps$. We discuss the integration of such detectors into quantum dot loaded, semiconductor ridge waveguides, resulting in the on-chip, time-resolved detection of quantum dot luminescence. Furthermore, a prototype quantum optical circuit is demonstrated that enabled the on-chip generation of resonance fluorescence from an individual InGaAs quantum dot, with a linewidth $< 15\,\mu eV$ displaced by $1\,mm$ from the superconducting detector on the very same semiconductor chip. Thus, all key components required for prototype quantum photonic circuits with sources, optical components and detectors on the same chip are reported.

**Main text:**

Semiconductors are ubiquitous in modern opto-electronics and quantum photonic devices and are also expected to play a major role in photonic quantum technologies [1]. In particular, semiconductor quantum dots have been shown to be near ideal sources of non-classical light [2] [3] that can be efficiently routed into nanoscale photonic circuits [4] [5] and even be used to manipulated the quantum state of single photons [4] [6] [7]. Besides the generation and routing of quantum light on a semiconductor chip [3] [8] [9], a further major requirement for realising integrated quantum devices is the on-chip detection of single photons with near unity quantum efficiency [10] and the integration of such detectors into nano-photonic circuits, consisting of waveguides, high-finesse nanocavities and beam splitters, opening the way toward integrated photon mediated quantum information technologies. Superconducting single photon detectors (SSPDs) based on Niobium nitride (NbN) nanowires combine high detection efficiency, low dark count rates, and picosecond timing resolution [11] [12], and can be integrated into prototype optical chips in a straightforward manner [13]. Altogether, the integration of SSPDs into nano-photonic circuits, their high detection efficiencies and ultra-fast response make them promising candidates for numerous quantum information applications [1], such as linear optical quantum computation [14], on-chip photonic quantum gates [15] [16], optical transistors [17] [18] or even quantum repeaters [19] [20].

In this article, we review our recent progress in the growth of high-quality superconducting NbN thin films with critical transition temperatures of $T_c > 12\ K$ and their patterning to realise highly efficient ($\eta > 20\%$) and ultra-fast ($< 72\ ps$) SSPDs on semiconducting GaAs substrates [21]. Furthermore, we demonstrate the integration of such SSPDs in quantum dot loaded photonic waveguides for on-chip time-resolved photoluminescence spectroscopy [22] [23]. Finally, we present a first prototype architecture of a fully integrated photonic nano-circuit, consisting of an individual InGaAs quantum dot, a GaAs/AlGaAs ridge waveguide and a waveguide-integrated SSPD, and demonstrate on-chip resonant fluorescence as a initial step towards on-chip quantum coherent information processing [24] [13].

Niobium nitride thin films were deposited on clean semiconducting gallium arsenide (GaAs) substrates using direct current (DC) reactive magnetron sputtering. This deposition technique has been shown to provide excellent control over the growth temperature $T_{gr}$, nitrogen partial pressure $P_{N_2}$ or sputtering rate $\Gamma_{NbN}$. High-quality superconducting NbN thin films have already been realized on various other substrates, for example sapphire [25], MgO [26], or glass [27], using this technique. The $\sim 5 \times 5\ mm^2$ GaAs substrates were mounted on a copper sample holder which can be heated up to $850\ °C$. After conditioning, the sputtering chamber is evacuated to a base pressure of $P_0 \leq 5 \cdot 10^{-6}\ mbar$ in order to reduce the partial pressure of residual gases that might become incorporated into the superconducting films as defects and, thus, potentially diminish their superconducting metrics such as critical temperature $T_c$ or transition width $\Delta T_c$. After conditioning, carefully measured quantities of process gases, Argon (Ar) and Nitrogen (N₂), are introduced into the growth chamber, controlled using mass-flow controllers. The mixture of these gases, i.e. their volume ratio primarily controls the Nb:N stoichiometry and, therefore, influences the resulting crystal structure of the NbN films [28] [21]. After reaching stable conditions with respect to gas flow and pressure, the plasma is started by applying a negative DC voltage of $\sim 350\ V$, thus, accelerating Ar⁺ ions towards the sputter target. During this pre-sputtering for $\sim 15\ min$, the top-oxide layer of the Nb-target is removed and a total chamber pressure of typically $P_{tot} = 4.4 \cdot 10^{-3}\ mbar$ is reached. After opening the shutter in front of the GaAs substrates the actual thin film deposition starts, whereby Nb atoms hit the sample surface and subsequently form chemical bonds with N₂ and, thus, arrange into crystal grains with a size and crystal structure that depends strongly on the chosen sputtering parameters. Typical film thicknesses investigated are in the range of $5 - 25\ nm$, determined by atomic force microscopy performed post growth.

The superconducting NbN thin films are electrically characterized employing temperature dependent transport measurements using a four-point geometry [29]. A typical result of the film resistance $R$ as a function of temperature $T$ is shown by the blue curve in Figure 1 (a) for a NbN film with thickness $d_{NbN} = 26.0 \pm 0.5\ nm$. We observe a gradual decrease of $R$ with decreasing $T$ with a drop to a value close to $0\Omega$ for $T \leq 11.8\ K$, reflecting the normal to superconducting phase transition. From such measurements, we deduce the two main figures of merit of the NbN films, namely the critical temperature $T_c$ and transition width $\Delta T_c$. Both quantities are extracted from the film resistivity measurements according the analysis in Ref. [30]. Based on the normalized numerical derivative $\Delta R/\Delta T$ as shown by the red curve in Figure 1(a), we define the temperature $T_{normal}$ at which the film remains just above the transition but still in its normal conducting state, and the $T_{super}$ where the film becomes superconducting. We define now the average normal state resistivity $\bar{R}$ as the arithmetic mean of the resistance between $T_{normal}$ and $T_{super} + 2\ K$. As shown in Figure 1(a), we can define now $T_{lower}$ and $T_{upper}$ as the boundaries of the $10\% - 90\%$ transition as $R(T_{lower}) = 0.1\ \bar{R}$ and $R(T_{upper}) = 0.9\ \bar{R}$, respectively. This provides an objective definition of the critical temperature $T_c \equiv 1/2(T_{upper} + T_{lower})$ and the transition width $\Delta T_c \equiv T_{upper} - T_{lower}$, giving rise to $T_c = 12.1 \pm 0.2\ K$ and $\Delta T_c = 0.3 \pm 0.05 K$ for the $d_{NbN} = 26 \pm 0.5 nm$ film presented in Figure 1(a).

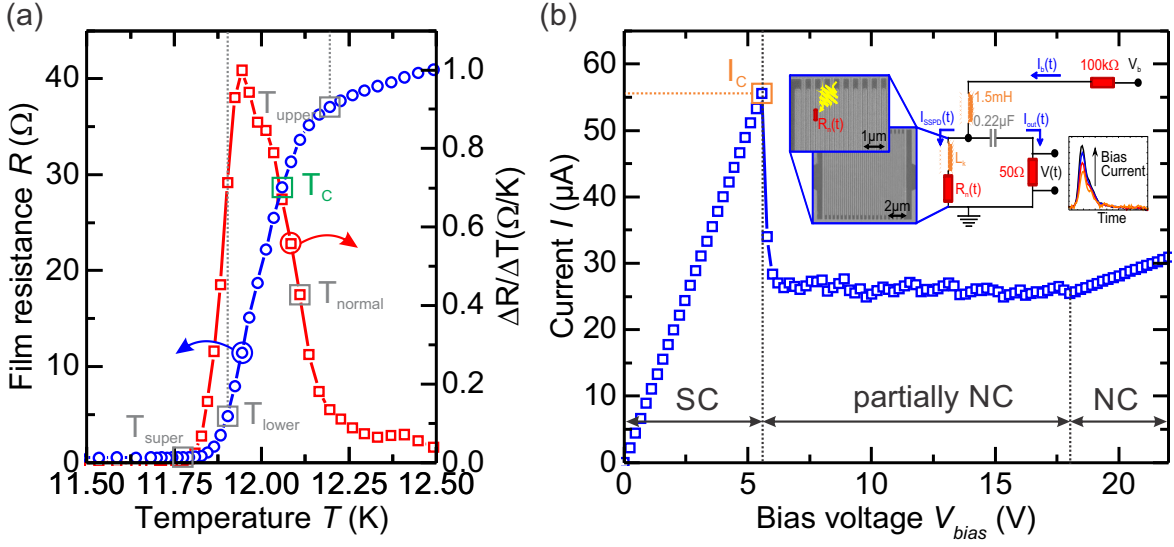

Figure 1 **(a)** Measured film resistance $R$ and derivative $\Delta R/\Delta T$ as a function of temperature $T$ for a NbN thin film with thickness $d_{NbN} = 26.0 \pm 0.5\ nm$ in red and blue, respectively. **(b)** Measured current $I$ as a function of applied bias voltage $V_{bias}$ for a NbN SSPD with thickness $d_{NbN} = 20.0 \pm 0.5\ nm$ as a temperature $T = 4\ K$. Inset: Schematic of electrical circuit used for SSPD electro-optical characterisation.

In a detailed study [21] we investigated NbN films with thicknesses $d_{NbN} = 4 - 20\ nm$ as a function growth temperature $T_{gr}$ and nitrogen partial pressure $P_{N_2}$. On such optimized films we established SSPDs employing a combination of electron beam lithography with a negative tone resist and reactive ion etching using a $SF_6/C_4F_8$ plasma [22] [24] . Nanowire SSPDs were arranged in a meander-type fashion with typically $34\times$ loops, nanowire widths $80 \pm 10\ nm$, nanowire separation $170 \pm 10\ nm$ and a total length of $23\ \mu m$ as shown in the scanning electron microscope images in the inset of Figure 1(b). In order to determine the the performance metrics of our SSPDs, we performed opto-electrical characterization measurements in a low-temperature ($T = 4.3 \pm 0.1\ K$) microwave probe station where individual devices can be contacted using mechanically adjustable microwave probes. A schematic circuit diagram of the setup used for probing the optical response of the detectors is shown in the inset of Figure 1(b). Hereby, the SSPD is operated just below the critical current $I_c$ using a bias-tee ($1\ mH$ inductance and $0.22 \mu F$ capacitance) connected to a $100\ k\Omega$ resistor and a voltage source $V_b$. Here, we assume the illuminated SSPD to be equivalent to an inductor $L_k$ with a time-dependent resistor $R(t)$.

Photon absorption events will result in locally formed normal conducting regions with $R_n$ and, thus, redirect the current onto the output-capacitor, eventually leading to a measureable voltage pulse $V(t)$. In the main panel of Figure 1(b), we present a typical current-voltage characteristic of a NbN SSPD with a thickness $d_{NbN} = 20 \pm 0.5 nm$ at $T = 4.2\,K$. The measured current $I$ increases linearly with a slope of $1/100\,k\Omega$, determined by the bias resistor, up to a maximum current of $I_c = 55.6\,\mu A$, as indicated by the orange dashed line. For voltages exceeding $\sim 5.6\,V$, $I$ drops to a value of $26.7 \pm 0.9\,\mu A$ and remains constant up to $V_{bias} = 18\,V$. For voltages larger $18\,V$, the current shows an ohmic increase corresponding to the SSPD resistance of $637\,k\Omega$. In general, we can divide the current-voltage curve shown in Figure 1(b) into three distinct regimes; (i) For $V_{bias} \leq 5.6\,V$ the device is in its superconducting state, (ii) for $5.7 \leq V_{bias} \leq 18\,V$ the SSPD is in a partially normal conducting state after the applied current $I$ exceeded $I_c$ and, thereby, caused a phase transitions from the superconducting to the normal conducting state for the thinnest, i.e. most constricted, nanowires [31] and (iii) for $V_{bias} \geq 18\,V$ the complete device has switched to a normal conducting state. The opto-electrical characterisation of our SSPDs enables us to reliably determine the critical current $I_c$ and on test the homogeneity of the lithographically defined superconducting nanowires.

The optical properties of SSPDs were investigated by measuring the time evolution, amplitude and lateral extension of the thermal hotspot formed in thin and thick devices upon single photon registration [21]. Having identified optimized SSPD designs, we continued to determine the detection efficiency $\eta$ and dark count rate $DC$, as the two most important figures of merit. Both quantities were measured with a setup as schematically depicted in the inset of Figure 1(b), with two low-noise $30\,dB$ high-bandwidth amplifiers connected to a $350\ MHz$ frequency counter. The SSPDs were illuminated by a carefully calibrated laser beam with a spot diameter $d_{beam} = 5.5 \pm 0.1\ mm$ whilst recording the recorded number of events per second as shown in Figure 2 (a). Here, two sets of data are presented for SSPDs with film thicknesses $d_{NbN} = 4 \pm 0.5\ nm$ and $d_{NbN} = 22 \pm 0.5\ nm$ as filled and open symbols, respectively. During each measurement, the SSPD under test is operated at a fixed current ratio $I/I_c$ as indicated by the numbers on the right of Figure 2 (a). Moreover, we indicated the corresponding dark count rates $DC$ by the dashed lines and according numbers on the left. For both SSPD thicknesses and all currents, we observe a clear linear increase of the detected single photon events as a function of the top-incident photon number in this double-logarithmic representation, strongly indicating that the detectors indeed work in the single photon regime [32] [33]. The dashed and solid curves in Figure 2 (a) are fits to the data, yielding exponents of $0.98 \pm 0.02$ and $1.09 \pm 0.03$ for the thin and thick SSPDs, respectively. The super-linear trend obtained for the thick detectors suggests that a small fraction of the detected counts is mostly likely arising from double photon events, in agreement with expectations for larger detector areas [32]. Moreover, for both batches of samples the dark count rate as well as the photon counts at constant photon flux increase exponentially with the current, in excellent agreements with previous reports on similar devices [33] [34].

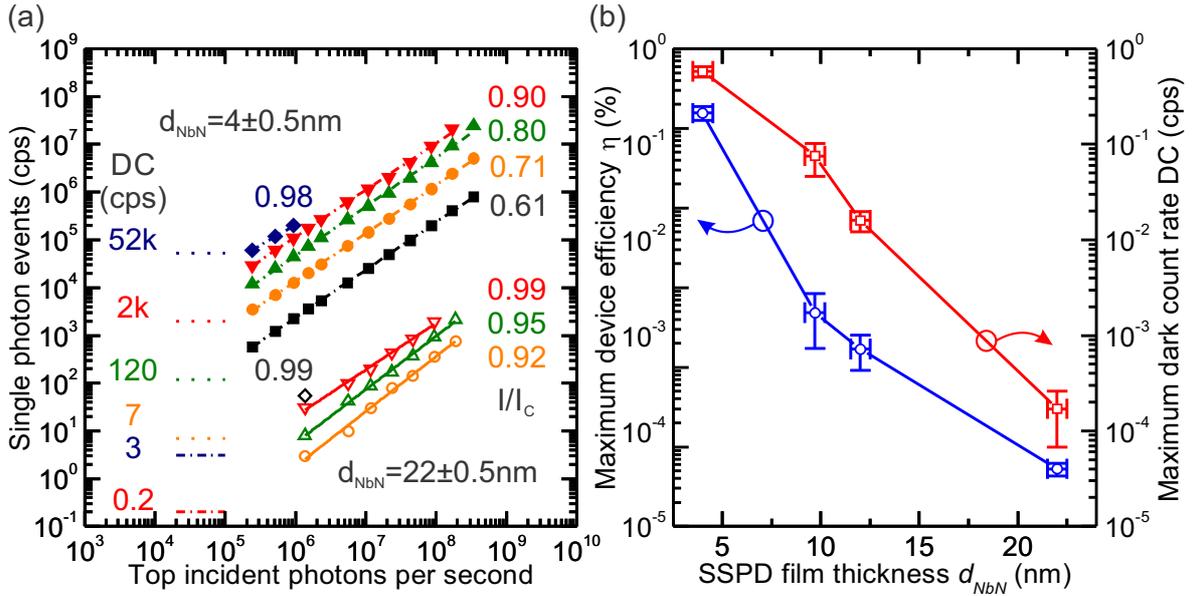

Figure 2 **(a)** Number of detected single photon events as a function of top-incident photon number for a $d_{NbN} = 4 \pm 0.5\,nm$ thin SSPD and a $d_{NbN} = 22 \pm 0.5\,nm$ thick SSPD in filled and open symbols, respectively. Numbers on the left and right give the according dark count rates $DC$ and currents $I$ with respect to $I_c$, respectively. **(b)** Maximum SSPD efficiency $\eta$ and corresponding dark count rate $DC$ as a function of SSPD film thickness $d_{NbN}$ in blue and red, respectively.

We continue to determined the maximum detection efficiency under top-illumination at a wavelength of $\lambda = 950 nm$ by dividing the number of detected counts by the number of incident photons at $0.98 \cdot I_{bias}/I_c$. Moreover, we maximised the incoming photon flux such than the detector did not permanently stay in a normal conducting state caused by a reduced critical current due to device heating [35]. The maximum device detection efficiency $\eta$ and the corresponding dark count rates $DC$ are shown in Figure 2(b) in blue and red, respectively. We obtain a $\eta = 21 \pm 2$ % for SSPDs with a thickness $d_{NbN} = 4 \pm 0.5\ nm$ at a dark count rate of $52000 \cdot 1/s$. This is in excellent agreement with values of 18.3 % for NbN SSPD detectors on GaAs at a illumination wavelength of $1.3 \mu m$ [36] as well as comparable studies on other substrates [37] [38] [39]. In strong contrast, we find for the thick SSPDs with $d_{NbN} = 22 \pm 0.5\ nm$ a much lower detection efficiency of $\eta = 0.004 \pm 0.001$ %, however, accompanied with an almost negligible dark count rate of only $3 \cdot 1/s$. The observed behaviour is fully consistent with the commonly accepted hotspot-model, where the absorption of a photon by a Cooper-pair leads to a small conducting region, the so-called hotspot, which eventually extends across the whole wire [40] [41] and furthermore in good accord with findings presented in Ref. [42].

In order to benefit from in-situ detection using SSPDs in functional semiconductor-based photonic circuits, a first step is their integration into prototype GaAs/AlGaAs ridge waveguides. To optimize the basic device characteristics, we applied numerical simulations to systematically study the propagating waveguide mode profiles, propagation and absorption losses and the dipole emission into the waveguide mode [43] using the commercial grade finite difference eigenmode (FDE) solver *Lumerical* [44]. The straight forward nanofabrication of SSPDs on top of ridge waveguides exploits the exponentially decaying field, outside the waveguide and inside the NbN detector and, thus, is based on the absorption of a constant fraction of the optical intensity per unit length. The feasibility of this intuitive approach has already been demonstrated for NbN-based SSPDs on both GaAs [45] and Si [10] waveguides, giving rise to up to 91 % single photon detection efficiencies for detectors of $40\ \mu m$ length.

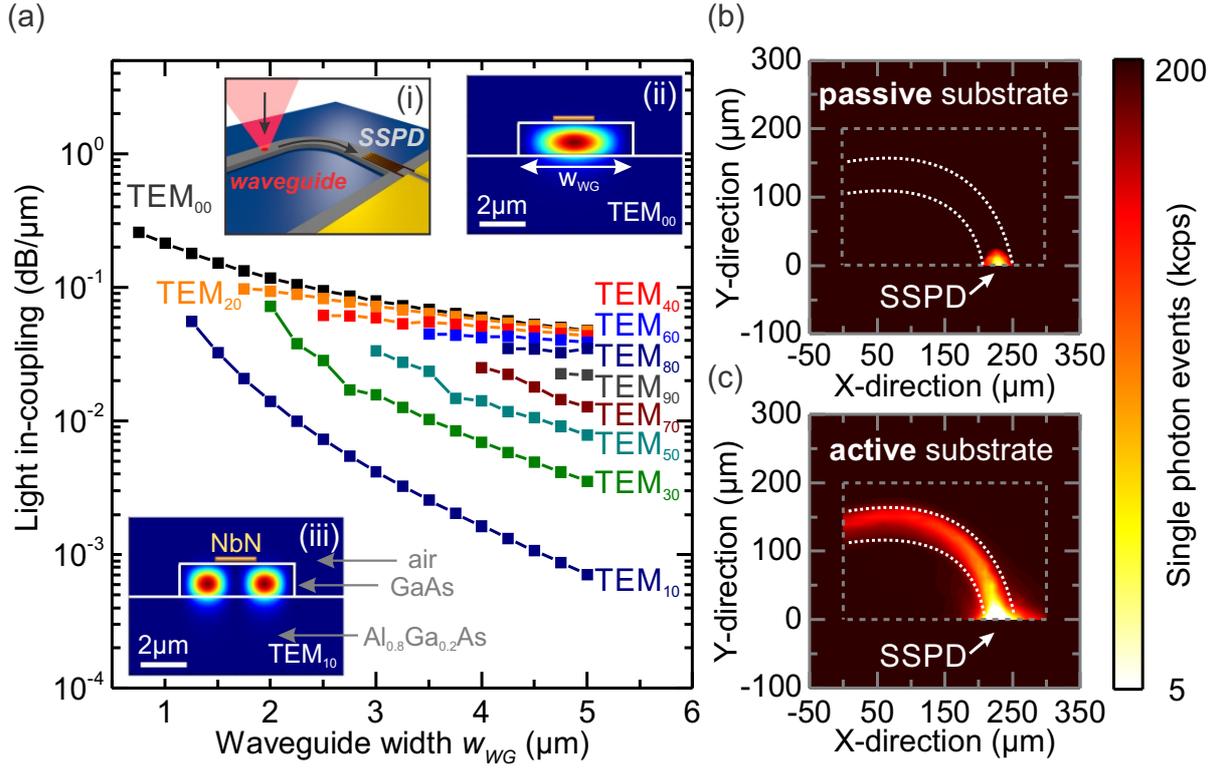

Figure 3 **(a)** Simulated efficiency of light incoupling from a waveguide into a waveguide-coupled SSPD for the first five even and odd modes as a function of waveguide width $w_{WG}$. The insets depict the mode distributions inside the waveguide for the $TEM_{00}$ and $TEM_{10}$ modes. **(b)** and **(c)** shows the detected single photon events as a function of excitation laser position for an energy $1.959\ eV$ for the passive and active substrate, respectively.

As an example, we show in Figure 3 (a) the simulated incoupling of light with an emission energy of $1.305 \pm 0.073\ eV$, corresponding to typical emission of InAs quantum dots, from the $Al_{0.8}Ga_{0.2}As/GaAs$ ridge waveguide into the detector (cp. inset (i)) for the first five even (TEM$_{x0}$ with $x = 0,2,4,6,8$) and odd (TEM$_{x0}$ with $x = 1,3,5,7,9$) modes as a function of the waveguide widths $w_{WG}$. The insets (ii) and (iii) of Figure 3 (a) represent cross-sections of the simulated structures and the mode profiles for the TEM$_{00}$ and TEM$_{10}$ modes. The SSPD in this simulation consists of 6 NbN nanowires of $100\ nm$ width. As expected, we obtain an increased number of guided modes when increasing $w_{WG}$ from $0.75\ \mu m$ to $5\ \mu m$ from just the first even mode TEM$_{00}$ to a total of 10 modes for the widest simulated waveguide. This trend is accompanied by a gradually decreasing light-incoupling for the TEM$_{00}$ mode from ~ $0.26\ dB/\mu m$ for $w_{WG} = 0.75\ \mu m$ by a factor of $5\times$ for the largest waveguide width studied $w_{WG} = 5\ \mu m$. This pronounced decrease in light-incoupling efficiency is attributed to a decreased overlap of the light field and the active NbN detector region, due to the fact that more substantially more light is guided off-centre for wider waveguides. This effect is even more pronounced for odd modes due to the existence of an anti-node at the waveguide centre. Whilst we observe a pronounced reduction of the light-incoupling by a factor of ~ $3.2\times$ for the lowest order odd mode TEM$_{10}$ when compared to the lowest order even mode TEM$_{00}$ at $w_{WG} = 1.25\ \mu m$, which further increases by a factor of ~ $67$ as the waveguide width increases to $w_{WG} = 5\ \mu m$, we also observe that this deviation lessens when comparing higher order odd and even modes, such as for example TEM$_{80}$ and TEM$_{90}$. This effect is explained by the increased number of nodes at the waveguide centre for higher order odd modes, which counteracts the initially lowered light-incoupling efficiency due to the node at the waveguide centre. The simulations in Figure 3 (a) suggest the use of single-mode waveguides with $w_{wg} < 1\ \mu m$ for maximum coupling efficiency between waveguides and SSPD, however, we note that additional simulations as a function of the filling factor $\eta = w_{NbN}/w_{WG}$ between active NbN material and waveguide width allow further optimization of the light-incoupling efficiency [43].

In order to demonstrate the functionality of waveguide-integrated SSPDs we fabricated two samples with bent waveguides as shown in inset (i) of Figure 3 (a), containing *no* quantum dots and an *ensemble* of quantum dots, which we term in the following *passive* and *active* substrate, respectively. We performed a low-temperature ($T = 4.2\ K$) optical experiment, where we spatially scanned a $\sim 5\ \mu m$ diameter excitation spot across the waveguide-detector area whilst simultaneously detecting the on-chip luminescence using the waveguide-integrated SSPD. The result for the passive substrate under optical illumination with light emitted of energy $1.959\ eV$ and an excitation power density of $4.2 \pm 0.4\ W/cm^2$ is shown in Figure 3 (b), where we observe only signal when directly illuminating the detector area. In strong contrast, we can clearly identify the bent waveguide structure when repeating the same measurement on the active substrate, as shown in Figure 3 (c). The strong signal obtained when exciting the waveguide structure far away from the detector region on the active substrate and the absence of signal in a similar experiment on the passive substrate, unambiguously proves on-chip detection of quantum dot luminescence using waveguide-integrated SSPDs [22].

In Figure 4 (a), we present the spectrally integrated quantum dot photoluminescence as a function of excitation energy for off-chip detection using a standard confocal microscope and on-chip detection using the waveguide-integrated SSPD in blue and red, respectively. We observe a reduction of the measured photoluminescence signal by $\sim 10^3 - 10^4 \times$ for both confocal and on-chip detection geometries as the excitation energy is tuned below $\sim 1.5\ eV$, which corresponds to the GaAs band gap at cryogenic temperatures. The observed trend is attributed to the strongly reduced absorption strength of the quantum dots when the excitation energy is tuned from above the GaAs band gap via the two-dimensional InGaAs wetting layer states ($E_{WL} \sim 1.4\ eV$) and into the s-shell transitions of the quantum dots ($E_{QD}^{s-shell} \sim 1.3\ eV$). Moreover, we also show in Figure 4 (a) the measured SSPD detection efficiency in green. In strong contrast to the strongly reduced quantum dot luminescence signal, we obtain for the SSPD detection efficiency only a weak reduction of $\sim 2.5 \times$ with decreasing excitation energy. This weak reduction of the SSPD detection efficiency is related to the size- and energy-dependence of the photon-inducted hotspots [32]; Thereby, detected photons with larger energy create larger hotspots and, thus, result in an enhanced probability to completely switch the nanowire in a normal conducting state. This has recently been demonstrated by energy-dependent measurements of the detector efficiency [46]. Since the propagation losses in the considered multi-mode waveguides are rather insensitive with respect to detection wavelength, our observations unambiguously show that the on-chip detected events are stemming from quantum dot emission into guided waveguide modes.

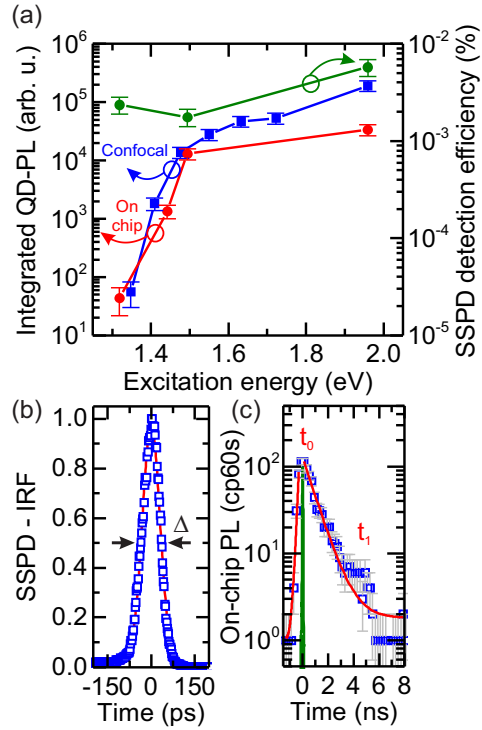

Figure 4 **(a)** Left axis: Integrated on-chip and confocal quantum dot photoluminescence as a function of excitation energy in red and blue, respectively. Right axis: Corresponding SSPD detection efficiency. **(b)** Instrument response function of a typical SSPD for a NbN film thickness of $d_{NbN} = 10 \pm 0.5\,nm$, operated at $T = 4.2\,K$ and $I_{bias} = 0.95 \cdot I_c$. The red line represents a Gaussian fit tot he data. (b) On-chip detected quantum dot photoluminescence as a function of the decay time. The red curves are exponential fits to the data. The green curve is the corresponding instrument response function.

In addition to on-chip photon detection, SSPDs also exhibit extremely high timing precision, which typically outperforms even state-of-the-art semiconductor-based single photon avalanche photodiodes [47]. Here, recent experiments have demonstrated for SSPDs photo-detection timing jitters as low as $18\,ps$ [38] [10]. In Figure 4 (b), we present the measured instrument response function of a SSPD with $d_{NbN} = 10 \pm 0.5\,nm$ under normal incidence illumination, operated at $T = 4.2\,K$ and $I_{bias} = 0.95 \cdot I_c$. Here, we illuminated the detector with a ps-pulsed laser diode emitting $\sim 60\,ps$ pulses at an energy of $1.302\,eV$ and a repition rate of $80\,MHz$. The red solid line in Figure 4 (b) represents a Gaussian fit to the data and yields a timing resolutions given by the full width at half maximum of $\Delta = 72 \pm 2\,ps$. Although, the value for the timing resolution is most likely limited by the $\sim 60\,ps$ pulse duration of the used pulsed laser diode, we note that it is much faster than typical quantum dot exciton lifetimes of $\sim 1\,ns$ [48] and, thus, enables us to perform on-chip time-resolved quantum dot spectroscopy.

In Figure 4 (c), we present typical on-chip, time-resolved photoluminescence data of a quantum dot ensemble located $\sim 0.5\,mm$ displaced from the SSPD, close to the remote end of a bended waveguide structure as schematically shown in inset (i) of Figure 3 (a). Exponential fits to the rising and falling edge of the photoluminescence intensity as a function of time are shown in red and yield a rise time $t_0 = 136 \pm 21\,ns$ and a fall time $t_1 = 0.95 \pm 0.03\,ns$, respectively. Here, the measured $t_1$-time is in excellent agreement with values of the spontaneous emission liftetimes of InGaAs quantum dots [48] [49]. The rise time reflects the timescale for charge carrier capture and thermalisation processes to the lowest unoccupied energy states of the quantum dot and the surprisingly slow rise-time clearly suggests the presence of a phonon bottleneck [50]. In our experiments, the excitation peak power density $P_{exc} \sim 25\ W/cm^2$ is kept sufficiently low in order to stay in the single exciton limit and, therefore, significant free carrier populations are not present in the wetting layer or the GaAs matrix. However, such free carriers are typically required to provide faster carrier capture and intra-dot carrier relaxation dynamics [50]. A detailed study of the carrier relaxation dynamics in single InGaAs quantum dots probed by on-chip integrated SSPDs can be found in Ref. [23].

After demonstrating the basic working principle of monolithically integrated quantum light sources [2], photonic quantum channels [1] and ultra-fast and highly efficient single photon detectors, we developed a first prototype of an integrated quantum optical circuit [13]. In Figure 5 (a), we show a typical light microscope image of a ready-fabricated sample after SSPD definition, waveguide nano-fabrication and bonding. The sample consists of a $2.6\ mm$ long, straight GaAs/AlGaAs multimode ridge waveguide into which a single layer of low-density, optically active self-assembled InGaAs quantum dots has been embedded at its midpoint. At the leftmost end of the waveguide, two SSPDs have been defined as shown by the scanning electron microscopy images in the inset of Figure 5 (a), which have been buried below a teflon-aluminium-teflon multilayer in order to reduce residual laser stray light. Moreover, a $1\ \mu m$ thick Si abosorber layer has been grown on the backside of the $350\ \mu m$ thick GaAs substrate in order to suppress back-reflected laser light from the sample backside [13]. A typical photoluminescence spectrum of a single InGaAs quantum dot under excitation at $E_{exc} = 1347.7\ eV$ via the two-dimensional wetting layer is shown in Figure 5 (b), exhibiting sharp and well-distinct emission lines, which are attributed to quantum dot emission [51]. For example the emission labelled $X_3$ exhibits a clear linear dependence as a function of the excitation power density as shown in the inset of Figure 5 (b), which is indicative for single excitonic transitions.

In order to study the $X_3$-emission in an on-chip detection geometry via the established SSPDs, we need to resonantly excite this quantum dot transition whilst suppressing scattered laser light, due to the lack of spectral filtering. Hereby, we tuned a single frequency laser in steps of $\delta E = 1\ \mu eV$ across the $X_3$-transition and record simultaneously the on-chip detected time-resolved fluorescence signal. Beside the already mentioned Si absorber layer, temporal filtering has been employed in order to efficiently suppress the laser [13]. Moreover, it has proven to be crucial to employ an additional, weak non-resonant laser with an excitation energy above the GaAs band gap ($E_{gate} = 1.534\ eV$, $P_{gate} = 0.06 \pm 0.02\ W/cm^2$), which continuously creates free charge carriers in the vicinity of the quantum dot under study and, therefore, fills up charge traps and stabilizes the otherwise fluctuating electrostatic charge environment [52] [53]. In Figure 5 (c), we show the on-chip recorded resonance fluorescence signal as a function of excitation laser detuning $\Delta E_{exc} = E_{exc} - E_{X_3}$ for gating laser power densities $P_{gate,0} = 0.06 \pm 0.02\ W/cm^2$ and $P_{gate,1} = 0.1 \cdot P_{gate,0}$ in the upper and lower panel, respectively. Here, $E_{exc}$ and $E_{X_3}$ denote the excitation laser energy and the energy of the $X_3$-transition, respectively. Both data sets exhibit a clear maximum of the resonance fluorescence signal $E_{exc} = E_{X_3}$. Fitting the data with Lorentzians, we obtain line width $\Delta_{FWHM}$ of $13.0 \pm 4.3\ \mu eV$ and $24.5 \pm 3.2\ \mu eV$ for the data set in the upper and lower panel of Figure 5 (c), respectively. For a reduced gating power density $P_{gate,1} = 0.1 \cdot$ $P_{gate,0}$, we observe a pronounced broadening of the line width, indicative for the onset of spectral wandering due to the more strongly fluctuating charge environment in the the vicinity of the quantum dot. However, we note that the minimum line width observed in such on-chip detected resonance fluorescence experiments is on the order of $\sim 14\ \mu eV$, in good agreement with values recently reported for inhomogeneously broadened quantum dot emission [53]. In order to demonstrate coherent Rabi oscillations [54] or coherently scattered single photons [55], further improvements in laser stray light suppression and the control of the local charge environment are required.

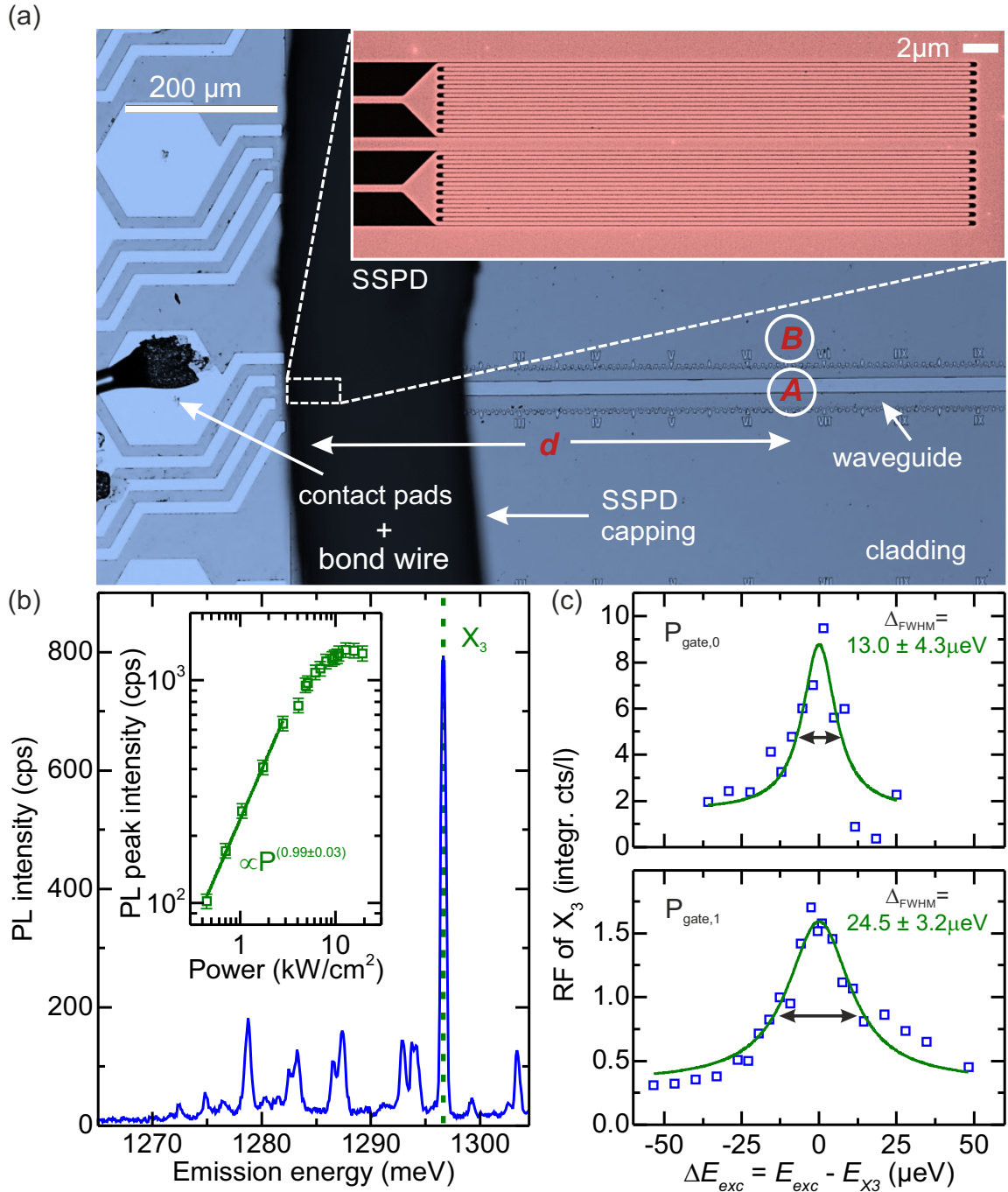

Figure 5 **(a)** Optical light microscope image of the prototype on-chip photonic circuit, consisting of a > $1\,mm$ long waveguide and two SSPDs integrated at one waveguide end. **(b)** Off-chip recorded photoluminescence spectrum of a single InGaAs quantum dots. Inset shows the photoluminescence intensity of the $X_3$-transition as a function of excitation power density. **(c)** Integrated resonance fluorescence signal of $X_3$ as a function of excitation laser detuning $\Delta E_{exc}$ for gating laser power $P_{gate,0}$ and $P_{gate,1}$ in the top and bottom panel, respectively.

In summary, we presented recent achievements in the NbN film growth, nanofabrication and integration of high-quality, high-efficiency and ultra-fast superconducting nanowire single photon detectors on optically active semiconducting GaAs substrates. Such detectors reach $> 20\ \%$ detection efficiencies for film thicknesses $\sim 4\ nm$, timing resolutions $< 72\ ps$ and are capable to be integrated in nano-photonic hardware such as the demonstrates GaAs/AlGaAs ridge waveguides. We further showed efficient on-chip generation and routing of quantum light emitted from embedded single InGaAs quantum dots and the subsequent time-resolved on-chip detection using the nano-fabricated SSPDs. Moreover, we addressed in a proof-of-concept experiment all necessary ingredients for future, fully-integrated quantum photonic circuits, by demonstrating the efficient generation, routing and on-chip detection of resonance fluorescence from a single semiconductor quantum dot, giving rise to line-width of $< 15\ \mu eV$; a key element for the exploitation of individual light quanta in future quantum photonic information technologies.

## Acknowledgements:

We gratefully acknowledge D. Sahin and A. Fiore (TU Eindhoven), K. Berggren and F. Najafi (MIT), and R. Hadfield (University of Glasgow) for useful discussions and the financial support from BMBF via QuaHL-Rep, project number 01BQ1036, Q.com via project number 16KIS0110, the EU via the integrated project SOLID and the DFG via SFB 631-B3.

## List of References:

[1] J. L. O'Brien, A. Furusawa and J. Vuckovic, "Photonic quantum technologies," *Nature Photonics,* vol. 3, p. 687, 2009.

[2] A. J. Shields, "Semiconductor quantum light sources," *Nature Photonics,* vol. 1, p. 215, 2007.

[3] A. Laucht, S. Pütz, T. Günther, N. Hauke, R. Saive, S. Frédérick, M. Bichler, M.-C. Amann, A. W. Holleitner, M. Kaniber and J. J. Finley, "A Waveguide-Coupled On-Chip Single-Photon Source," *Phys. Rev. X,* vol. 2, p. 011014, 2012.

[4] A. Faraon, I. Fushman, D. Englund, N. Stoltz, P. Petroff and V. Vuckovic, "Coherent generation of non-classical light on a chip via photon-induced tunnelling and blockade," *Nature Physics,* vol. 4, p. 859, 2008.

[5] A. Politi, J. Matthews, M. G. Thompson and J. L. O'Brien, "Integrated quantum photonics," *IEEE J. Quantum Electron,* vol. 15, p. 1673, 2009.

[6] J. C. F. Matthews, A. Politi, A. Stefanov and J. L. O'Brien, "Manipulation of multiphoton entanglement in waveguide quantum circuits," *Nature Photonics,* vol. 3, p. 346, 2009.

[7] P. Lodahl, S. Mahmoodian and S. Stobbe, "Interfacing single photons and single quantum dots with photonic nanostructures," *Rev. Mod. Phys.,* vol. 87, p. 347, 2015.

[8] A. Schwagmann, S. Kalliakos, I. Farrer, J. P. Griffiths, G. A. C. Jones, D. A. Ritchie and A. J. Shields, "On-chip single photon emissio from an integrated semiconductor quantum dot into a photonic crystal waveguide," *Appl. Phys. Lett.,* vol. 99, p. 261108, 2011.

[9] T. B. Hoang, J. Beetz, L. Midolo, M. Skacel, M. Lermer, M. Kamp, S. Höfling, L. Balet, N. Chauvin and A. Fiore, "Enhanced spontaneous emission from quantum dots in short photonic crystal waveguides," *Appl. Phys. Lett.,* vol. 100, p. 061122, 2012.

[10] W. H. P. Pernice, C. Schuck, O. Minaeva, M. Li, G. N. Goltsman, A. V. Sergienko and H. X. Than, "High-speed and high-efficiency travelling wave single-photon detectors embedded in nanophotonic circuits," *Nature Communications,* vol. 3, p. 1325, 2012.

[11] G. N. Gol'tsman, O. Okunev, G. Chulkova, A. Lipatov, A. Semenov, K. Smirnov, B. Voronov, A. Dzardanov, C. Williams and R. Sobolewski, "Picosecond superconducting single-photon optical detector," *Appl. Phys. Lett.,* vol. 79, p. 705, 2001.

[12] F. Najafi, F. Marsili, E. Dauler, R. J. Molnar and K. K. Berggren, "Timing performance of 30-nm-wide superconducting nanowire avalanch photodetectors," *Appl. Phys. Lett.,* vol. 100, p. 152 602, 2012.

[13] G. Reithmaier, M. Kaniber, F. Flassig, S. Lichtmannecker, K. Müller, A. Andrejew, J. Vuckovic, R. Gross and J. J. Finley, "On-Chip Generation, Routing, and Detection of Resonance fluorescence," *Nano Letters,* vol. 15, p. 5208, 2015.

[14] E. Knill, R. Laflamme and G. J. Milburn, "A scheme for efficient quantum computation with linear optics," *Nature,* vol. 409, p. 46, 2001.

[15] D. E. Chang, A. S. Sorensen, E. A. Demler and M. D. Lukin, "A single-photon transitor using nanoscale surface plasmons," *Nature Physics,* vol. 3, p. 807, 2007.

[16] J. Hwang, M. Pototschnig, R. Lettow, G. Zumofen, A. Renn, G. S. and V. Sandoghdar, "A single-molecule optical transistor," *Nature,* vol. 460, p. 76, 2009.

[17] D. Tiarks, S. Baur, K. Schneider, S. Dürr and G. Rempe, "Single-Photon Transistor Using a Förster Resonance," *Phys. Rev. Lett.,* vol. 113, p. 053602, 2014.

[18] H. Gorniaczyk, C. Tresp, J. Schmidt, H. Fedder and S. Hofferberth, "Single-Photon Transistor Mediated by Interstate Rydberg Interactions," *Phys. Rev. Lett.,* vol. 113, p. 053601, 2014.

[19] Z.-S. Yuan, Y.-A. Chen, B. Zhao, S. Chen, J. Schmiedmayer and J.-W. Pan, "Experimental demonstration of a BDCZ quantum repeater node," *Nature,* vol. 454, p. 1098, 2008.

[20] H.-J. Briegel, W. Dürr, J. I. Cirac and P. Zoller, "Quantum Repeaters: The Role of Imperfect Local Operations in Quantum Communication," *Phys. Rev. Lett. ,* vol. 81, p. 5932, 1998.

[21] G. Reithmaier, J. Senf, S. Lichtmannecker, T. Reichert, F. Flassig, A. Voss, R. Gross and J. J. Finley, "Optimisation of NbN thin films on GaAs substrates for in-situ single photon detection in structured photonic devices," *J. Appl. Phys.,* vol. 113, p. 143507, 2013.

[22] G. Reithmaier, S. Lichtmannecker, T. Reichert, P. Hasch, K. Müller, M. Bichler, R. Gross and J. J. Finley, "On-chip time resolved detection of quantum dot emission using integrated superconducting single photon detectors," *Scientific Reports,* vol. 3, p. 1901, 2013.

[23] G. Reithmaier, F. Flassig, P. Hasch, S. Lichtmannecker, K. Müller, J. Vuckovic, R. Gross, M. Kaniber and J. J. Finley, "A carrier relaxation bottleneck probed in single InGaAs quantum dots using integrated superconducting single photon detectors," *Appl. Phys. Lett.,* vol. 105, p. 081107, 2014.

[24] F. Flassig, M. Kaniber, G. Reithmaier, K. Müller, A. Andrejew, R. Gross, J. Vuckovic and J. J. Finley, "Towards on-chip generation, routing and detection of non-classical light," *Proc. of SPIE,* vol. 9373, p. 937305, 2015.

[25] J. Villegirr, N. Hadacek, S. Monso, B. Delnet, A. Roussy, P. Febvre, G. Lamura and J. Laval, "NbN multilayer technology on R-plane sapphire," *IEEE Transactions on Applied Superconductivity,* vol. 11, p. 68, 2001.

[26] F. Marsili, D. Bitauld, A. Fiore, A. Gaggero, F. Mattioli, R. Leoni, M. Benkahoul and F. Lévy, "High efficiency NbN nanowire superconducting single photon detectors fabricated on MgO substrates from a low temperature process," *Optics Express,* vol. 16, p. 3191, 2008.

[27] W. N. Maung, D. P. Butler and C. A. Huang, "Fabrication of NbN thin films by reactive sputtering," *J. Vac. Sci. Technol. A: Vacuum, Surfaces, and Films,* vol. 11, p. 615, 1993.

[28] M. S. Wong, W. D. Sproul, X. Chu and S. A. Barnett, "Reative magentron sputter deposition of niobium nitride films," *J. Vac. Sci. Technol. A: Vacuum, Surfaces, and Films,* vol. 11, p. 1528, 1993.

[29] F. M. Smits, "Measurement of Sheet Resistivities with the Four-Point Probe," *Bell System Technical Journal,* vol. 37, p. 711, 1958.

[30] F. Marsili, A. Gaggero, L. H. Li, A. Surrente, R. Leoni, F. Lvy and A. Fiore, "High quality superconducting NbN thin films on GaAs," *Superconducting Science and Technology,* vol. 22, p. 095013, 2009.

[31] A. J. Kerman, E. A. Dauler, J. K. W. Yang, K. M. Rosfjord, V. Anant, K. K. Berggren and G. N. Goltsman, "Constriction-limited detection efficiency of superconducting nanowire single-photon detectors," *Appl. Phys. Lett.,* vol. 90, p. 101110, 2007.

[32] L. Maingault, M. Tarkhov, I. Floarya, A. Semenov, R. E. de Lamaestre, P. Cavalier, G. Gol'tsman, J.-P. Poizat and J.-C. Villéger, "Spectral dependency of superconducting single photon detectors," *J. Appl. Phys.,* vol. 107, p. 116103, 2010.

[33] A. Divochiy, F. Marsili, D. Bitauld, A. Gaggero, R. Leoni, F. Mattioli, A. Korneev, V. Seleznew, N. Kaurova, M. O., G. Gol'tsman, K. G. Lagoudakis, M. Benkhaoul, D. Lvy and A. Fiore, "Superconducting nanowire photonnumber-resolving detector at telecommunication wavelengths," *Nature Photonics,* vol. 2, p. 302, 2008.

[34] S. N. Dorenbos, P. Forn-Díaz, T. Fuse, A. H. Verbruggen, T. Zijlstra, T. M. Klapwijk and V. Zwiller, "Low gap superconducting single photon detectors for infrared sensitivity," *Appl. Phys. Lett.,* vol. 98, p. 251102, 2011.

[35] Z. Yan, A. Majedi and S. Safavi-Naeini, "Physical Modeling of Hot-Electron Superconducting Single-Photon Detectors," *IEEE Transactions on Applied Superconductivity,* vol. 17, p. 3789, 2007.

[36] A. Gaggero, S. J. Nejad, F. Marsili, F. Mattioli, R. Leoni, D. Bitauld, D. Sahin, G. J. Hamhuis, R. Nötzel, R. Sanjines and A. Fiore, "Nanowire superconducting single-photon detectors on GaAs for integrated quantum photonic applications," *Appl. Phys. Lett.,* vol. 97, p. 151108, 2010.

[37] A. Korneev, V. Matvienko, O. Minaeva, I. Milstnaya, I. Rubtsova, G. Chulkova, K. Smirnov, V. Voronov, G. Gol'tsman, W. Slysz, A. Pearlman, A. Verevkin and R. Sobolewski, "Quantum efficiency and noise equivalent power of nanostructured, NbN, single-photon detectors in the wavelength range from visible to infrared," *IEEE Transactions on Applied Superconductivity,* vol. 15, p. 571, 2005.

[38] G. N. Gol'tsman, A. Korneev, I. Rubtsova, I. Milstnaya, G. Chulkova, O. Minaeva, I. Smirnov, B. Voronov, W. Sysz, A. Pearlman, A. Verevkin and R. Sobolweski, "Ultra-fast superconducting single-photon detectors for near-infrared-wavelength quantum communications," *Physica Status Solidi (c),* vol. 2, p. 1480, 2005.

[39] G. Gol'tsman, O. Okunev, G. Chulkov, A. Lipatov, A. Dzardanov, K. Smirnov, A. Semenov, B. Voronov, C. Williams and R. Sobolewski, "Fabrication and properties of an ultrafast NbN hot-electron single-photon detector," *IEEE Transactions on Applied Superconductivity,* vol. 11, p. 574, 2001.

[40] W. J. Skocpol, M. R. Beasley and M. Tinkham, "Self-heating hotspots in superconducting thin-film microbridges," *J. Appl. Phys.,* vol. 45, p. 4054, 1974.

[41] A. D. Semenov, G. N. Goltsman and A. A. Korneev, "Quantum detection by current carrying superconducting film," *Physica C-superconductivity and Its Applications,* vol. 351, p. 439, 2001.

[42] M. Hofherr, D. Rall, K. Ilin, M. Siegel, A. Semenov, H.-W. Hubers and N. A. Gippius, "Intrinsic detection efficiency of superconducting nanowire single-photon detectors with different thicknesses," *J. Appl. Phys.,* vol. 108, p. 014507, 2010.

[43] G. M. Reithmaier, Superconducting detectors for semiconductor quantum photonics, München: Printy Digitaldruck, 2015.

[44] Lumerical Solutions, Inc., "http://www.lumerical.com/tcad-products/fdtd/," [Online].

[45] J. P. Spengers, A. Gaggero, D. Sahin, S. Jahanmirinejad, G. Frucci, F. Mattioli, R. Leoni, J. Beetz, M. Lermer, M. Kamp, S. Höfling, R. Sanjines and A. Fiore, "Waveguide superconducting single-photon detectors for integrated quantum photonic circuits," *Appl. Phys. Lett.,* vol. 99, p. 181110, 2011.

[46] F. Marsili, F. Bellei, F. Najafi, A. E. Dane, E. A. Dauler, R. J. Molnar and K. K. Berggren, "Efficient Single Photon Detection from 500 nm to 5μm Wavelength," *Nano Letters,* vol. 12, p. 4799, 2012.

[47] R. Hadfield, "Single-photon detectors for optical quantum information applications," *Nature Photonics,* vol. 3, p. 696, 2009.

[48] J. M. Gérard, O. Cabrol and B. Sermage, "InAs quantum boxes: Highly efficient radiative traps for light emitting devices on Si," *Appl. Phys. Lett.,* vol. 68, p. 3123, 1996.

[49] E. Viasnoff-Schwoob, C. Weisbuch, H. Bensity, S. Olivier, S. Varoutsis, I. Robert-Philip, R. Houdré and C. J. M. Smith, "Spontaneous Emission Enhancement of Quantum Dots in a Photonic Crystal Wire," *Phys. Rev. Lett.,* vol. 95, p. 183901, 2005.

[50] J. Urayama, T. B. Norris, J. Singh and P. Bhattacharaya, "Observation of Phonon Bottleneck in Quantum Dot Electronic Relaxation," *Phys. Rev. Lett.,* vol. 86, p. 4930, 2001.

[51] J. J. Finley, A. D. Ashmore, A. Lemaître, D. J. Mowbray, M. S. Skolnick, I. E. Itskevich, P. A. Maksym, M. Hopkinson and T. F. Krauss, "Charged and neutral exciton complexes in individual self-assembled In(Ga)As quantum dots," *Phys. Rev. B,* vol. 63, p. 073307, 2001.

[52] H. Nguyen, G. Sallen, C. Voisin, P. Roussignol, C. Diedrichs and G. Cassabois, "Optically Gated Resonant Emission of Single Quantum Dots," *Phys. Rev. Lett.,* vol. 108, p. 057401, 2012.

[53] M. N. Makhonin, J. E. Dixon, R. J. Coles, B. Royall, I. J. Luxmoore, E. Clarke, M. Hugues, M. S. Skolnick and A. M. Fox, "Waveguide Coupled Resonance Fluorescence from On-Chip Quantum Emitters," *Nano Letters,* vol. 14, p. 6997, 2014.

[54] A. Zrenner, E. Beham, S. Stufler, F. Findeis, M. Bichler and A. Abstreiter, "Coherent properties of a two-level system based on a quantum-dot photodiode," *Nature,* vol. 418, p. 612, 2002.

[55] C. Matthiesen, A. N. Vamivakas and M. Atatür, "Subnatural Linewidth Single Photons from a Quantum Dot," *Phys. Rev. Lett.,* vol. 108, p. 093602, 2012.